\documentclass[sn-nature]{sn-jnl}
\usepackage{lmodern}
\usepackage{graphicx}%
\usepackage{multirow}%
\usepackage{amsmath,amssymb,amsfonts}%
\usepackage{amsthm}%
\usepackage{mathrsfs}%
\usepackage[title]{appendix}%
\usepackage{xcolor}%
\usepackage{textcomp}%
\usepackage{manyfoot}%
\usepackage{booktabs}%
\usepackage{algorithm}%
\usepackage{algorithmicx}%
\usepackage{algpseudocode}%
\usepackage{listings}%
\usepackage[separate-uncertainty=true,multi-part-units=single]{siunitx}
\usepackage{caption}
\usepackage{tabularx}
\usepackage{bm}

\usepackage{pdfpages}

\DeclareSIUnit{\atomicunit}{au}

\newcommand{\ket}[1]{\vert #1 \rangle}

\newcommand{\bra}[1]{\langle #1 \vert}

\newcommand{\braket}[1]{\langle #1 \rangle}

\renewcommand{\theta}{\vartheta}

\newcommand\w[1]{\makebox[2.5em]{$#1$}}

\begin{document}

\title{Convex Hartree-Fock theory: A simple framework for ground state conical intersections}

\author[1]{\fnm{Federico} \sur{Rossi}}

\author*[1]{\fnm{Henrik} \sur{Koch}}
\email{henrik.koch@ntnu.no}

\affil[1]{\orgdiv{Department of Chemistry}, \orgname{Norwegian University of Science and Technology}, \orgaddress{\city{Trondheim}, \postcode{7491}, \country{Norway}}}

\abstract{Accurate modeling of conical intersections is crucial in nonadiabatic molecular dynamics, as these features govern processes such as radiationless transitions and photochemical reactions. Conventional electronic structure methods, including Hartree-Fock, density functional theory, and their time-dependent extensions, struggle in this regime. Due to their single reference nature and separate treatment of ground and excited states, they fail to capture ground state intersections. Multiconfigurational approaches overcome these limitations, but at a prohibitive computational cost. In this work, we propose a modified Hartree-Fock framework, referred to as Convex Hartree-Fock, that optimizes the reference within a tailored subspace by removing projections along selected Hessian eigenvectors. The ground and excited states are then obtained through subsequent Hamiltonian diagonalization. We validate the approach across several test cases and benchmark its performance against time-dependent Hartree-Fock within the Tamm-Dancoff approximation.} 

\maketitle

\section*{Introduction}\label{sec:intro} 
The electronic structure of molecules in regions where two or more states approach degeneracy presents one of the most complex challenges in quantum chemistry. In particular, the behavior of systems near ground state conical intersections demands a theoretical description capable of capturing pronounced multireference character. Traditional single-determinant frameworks like Hartree–Fock (HF) and density functional theory (DFT), despite their widespread use and computational efficiency, are fundamentally inadequate for this task. Both Hartree-Fock and conventional Kohn-Sham DFT rely on a mean-field approximation that implicitly assumes a dominant electronic configuration. This assumption breaks down in the vicinity of such conical intersections, where the ground state wave function becomes strongly mixed and cannot be described by a single Slater determinant. The results are discontinuities in the energies, incorrect topology of the potential energy surfaces, and convergence failures~\cite{gozem2014shape}. For DFT, the problem is further intensified by the fact that most exchange–correlation functionals are developed for a non-degenerate ground state, leading to unreliable predictions when this is no longer the case~\cite{capelle2007degenerate,filatov2013assessment}.

The stability of Hartree-Fock solutions has been studied in great detail previously~\cite{thouless2013quantum,linderberg1973,cizek1971}. Bifurcations of the solutions are well known to appear frequently on potential energy surfaces, for instance, the Coulson-Fischer point in the hydrogen molecule for an unrestricted Hartree-Fock wave function~\cite{coulson1949xxxiv, helgaker2013molecular}. The non-analytical behavior has been discussed by \v{C}\'{\i}\v{z}ek and Paldus~\cite{cizek1971} and was recently shown to be particularly pronounced in areas around ground state conical intersections~\cite{kjonstad2024understanding}. These limitations directly impact the accuracy of simulations in photochemistry and ultrafast spectroscopy. Due to the critical importance of ground state conical intersections, a range of computational solutions has been developed over the years to address these degeneracies within time-dependent mean-field methods. 
One example is spin-flip TDDFT~\cite{shao2003spin}, which redefines the reference state to access crossings between ground and excited states by allowing spin-changing excitations. Other approaches include modified exchange-correlation kernels to account for double excitation effects~\cite{maitra2004double}, and TDDFT-1D~\cite{teh2019simplest}, where specific excited configurations are incorporated into the eigenvalue problem. 
Configuration interaction-corrected DFT (CIC-TDA~\cite{li2014configuration}) restores the correct dimensionality of conical intersections by augmenting the Tamm–Dancoff approximation (TDA) with a perturbative configuration interaction scheme.
Further advances include orthogonality constrained DFT (OCDFT~\cite{evangelista2013orthogonality}), multiconfigurational short-range DFT~\cite{hedegaard2018multiconfigurational}, and ensemble-based DFT methods adapted to describe excited states with near-degeneracies (SI-SA-REKS~\cite{filatov2015spin}).
In addition, mode-following techniques~\cite{schmerwitz2023calculations} have been employed to directly locate excited state saddle points by following specific eigenvectors of the Hessian. Most recently, phase-space electronic structure theory~\cite{duston2025conical} reformulates electronic structure in terms of phase-space variables, offering new insights into conical intersections by explicitly incorporating electronic momentum.
These and other methods to address conical intersection problems have recently been reviewed by Matsika~\cite{matsika2021electronic}.

To our knowledge, no general procedure to resolve the aforementioned complications has been developed by correcting the wave function parameterization in Hartree-Fock theory. In this work, we introduce the Convex Hartree–Fock (CVX-HF) method, in which the stationarity conditions and the Hessian eigenvalue equations are solved self-consistently within a subspace where the Hessian is positive definite. This framework offers excellent convergence properties and introduces the necessary coupling elements in the Hamiltonian matrix to yield eigenstates that correctly capture conical intersections and the geometric phase effect. Since TDA is typically used to avoid complex excitation energies arising from the non-Hermitian TDHF eigenvalue problem~\cite{martinez2020}, we compare the performance of CVX-HF to TDHF with TDA in a variety of molecular systems.

\section*{Results}\label{sec:results_and_discussion}
\subsection*{Convex Hartree-Fock theory}
We start from a reference determinant $\ket{\Phi_0}$ constructed from orbitals obtained from the superposition of atomic densities (SAD)~\cite{almlof1982principles, van2006starting}. The Hartree-Fock wave function is parameterized in terms of a single global orbital rotation, such that 
\begin{align}
    \ket{\mathrm{HF}} = \exp(\sum_{ai} \kappa_{ai}E^-_{ai}) \ket{\Phi_0} ,\label{eq:Hartree-Fock}
\end{align}
where $E_{ai}^- =E_{ai} - E_{ia}$, $E_{ai}=\sum_\sigma a^\dagger_{a\sigma}a_{i\sigma}$, and $\kappa_{ai}$ are the orbital rotation parameters. We use $a,b$ to label virtual orbitals and $i,j$ for occupied orbitals.
The Hartree-Fock energy $\mathcal{E}_\text{HF}$ is obtained as the expectation value of the electronic Hamiltonian $H$,
\begin{align}
\mathcal{E}_\text{HF} = \bra{\mathrm{HF}}H\ket{\mathrm{HF}},
\end{align}
where we assume $\ket{\mathrm{HF}}$ is normalized.
When minimizing the energy with respect to the $\kappa_{ai}$ parameters, we need to calculate the electronic gradient and Hessian. These are conveniently obtained by considering the Hartree-Fock energy function
\begin{align}
\mathcal{E}(\bm\gamma) = \bra{\mathrm{HF}}\exp(-\gamma)H\exp(\gamma)\ket{\mathrm{HF}},
\end{align}
where $\exp(\gamma)$ is an orbital rotation. The electronic gradient and Hessian with respect to the $\gamma_{ai}$ parameters is given by
\begin{align}
   G^{(0)}_{ai}(\mathbf{0}) &= \frac{\partial \mathcal{E}}{\partial \gamma_{ai}}\Big|_{\bm\gamma=\mathbf{0}}=\bra{\mathrm{HF}}[H,E^-_{ai}]\ket{\mathrm{HF}}\\
   G^{(1)}_{ai,bj}(\mathbf{0}) &= \frac{\partial^2 \mathcal{E}}{\partial \gamma_{ai}\partial \gamma_{bj}}\Big|_{\bm\gamma=\mathbf{0}}=\frac{1}{2}P_{ai,bj}\bra{\mathrm{HF}}[[H,E^-_{ai}],E^-_{bj}]\ket{\mathrm{HF}}, \label{eq:Hessian_HF}
\end{align}
where the permutation operator is defined as $P_{ai,bj} A_{ai,bj} = A_{ai,bj} + A_{bj,ai}$.
The  $\kappa_{ai}$  parameters are determined by requiring the gradient to be zero and the Hessian to be positive definite, if possible. Once the Hartree-Fock state is determined, the excited states can be calculated using time-dependent Hartree-Fock (TDHF). The excitation energies $\omega$ are obtained from the non-Hermitian TDHF eigenvalue problem
\begin{align}
\begin{pmatrix}
\mathbf{A} & \mathbf{B} \\
\mathbf{B} & \mathbf{A} \\
\end{pmatrix}\begin{pmatrix}
\mathbf{X} \\
\mathbf{Y} \\
\end{pmatrix}=\omega\begin{pmatrix}
\mathbf{1} & \mathbf{0} \\
\mathbf{0} & -\mathbf{1} \\
\end{pmatrix}\begin{pmatrix}
\mathbf{X} \\
\mathbf{Y} \\
\end{pmatrix},
\label{eq:TD_theory}
\end{align}
where
\begin{align}
&A_{ia,jb} = \delta_{ij}\delta_{ab}(\varepsilon_a - \varepsilon_i)+2g_{aijb} - g_{abji}\\
&B_{ia,jb} = 2g_{aibj} - g_{ajbi},
\end{align} 
$g_{pqrs}$ are the two-electron integrals, and $\varepsilon_p$ are the orbital energies.
A comprehensive analysis of the stability of the TDHF eigenvalue problem has been given by J{\o}rgensen and Simons~\cite{jorgensen2012second}. In the Tamm-Dancoff approximation, the $\mathbf{B}$ matrix in Eq. \ref{eq:TD_theory} is discarded and the eigenvalue problem simply reads $\mathbf{A}\mathbf{x}=\omega \mathbf{x}$~\cite{hirata1999time}.

When optimizing the Hartree-Fock energy, we may encounter convergence problems or multiple solutions to the ground state equations ($\mathbf{G}^{(0)}=\mathbf{0}$). To analyze this situation, we Taylor expand the gradient condition as
\begin{align}
G^{(0)}_{ai}(\bm\gamma)=G^{(0)}_{ai}(\mathbf{0}) + \sum_{bj} G^{(1)}_{ai,bj} (\mathbf{0})\gamma_{bj}  + \ldots = 0.
\end{align}
Transforming to the eigenbasis of the Hessian, $\mathbf{G}^{(1)}\mathbf{r}_n=\lambda_n \mathbf{r}_n$, we obtain
\begin{align}
G^{(0)}_{n}(\bm\gamma)=G^{(0)}_{n}(\mathbf{0}) + \lambda_n \gamma_{n}  + \ldots = 0.
\end{align}
When approaching a conical intersection, the lowest positive eigenvalue $\omega_1$ in Eq. \ref{eq:TD_theory} becomes close to zero. And as zero-energy eigenvectors of Eq. \ref{eq:TD_theory} are also zero-energy eigenvectors of the Hessian~\cite{cui2013proper}, this implies that $\lambda_1$ will also approach zero. The first-order contribution is thus small, and the higher-order terms become important, giving rise to bifurcations in the energy.
We define a modified gradient $\tilde{\mathbf{G}}^{(0)}$, where the projection along the first eigenvector of the Hessian $\mathbf{r}_1$ is removed, i.e.
\begin{align}
\tilde{\mathbf{G}}^{(0)} = \mathbf{G}^{(0)} - \mathbf{r}_1 (\mathbf{r}_1^T \mathbf{G}^{(0)}).
\end{align}
This new gradient allows us to solve along all the other eigenvectors and to implicitly define an effective Hessian $\tilde{\mathbf{G}}^{(1)}$ that is positive definite, forcing the energy function to be convex. We can express the $\kappa$ operator in the basis of the Hessian eigenvectors
\begin{align}
\kappa = \sum_{ai} \kappa_{ai}E^-_{ai} = \sum_n \kappa_n R_n
\end{align}
where $R_n = \sum_{ai} r_{ai,n} E^-_{ai}\,.$
Since we employ a modified gradient, we remove the corresponding component from $\bm\kappa$
\begin{align}
    \tilde{\kappa}_{ai} = \kappa_{ai} - r_{ai,1} \sum_{bj}r_{bj,1} \kappa_{bj},
\end{align}
which guarantees $\tilde{\kappa}_1 = 0$.
The final set of convex Hartree-Fock equations is given by
\begin{align}
&\tilde{\mathbf{G}}^{(0)}=0\\
& \mathbf{G}^{(1)}\mathbf{r}_1 = \lambda_1 \mathbf{r}_1,
\end{align}
where $\tilde{\mathbf{G}}^{(0)} $ and $\mathbf{G}^{(1)}$ are evaluated with $\tilde{\bm\kappa}$.

A detailed description of the employed algorithm can be found in the Method section. Since the equations are solved in a specific subspace, we avoid convergence issues related to degeneracies. The projected component can be introduced in a final diagonalization of the Hamiltonian matrix, written in the basis of $\{ \ket{\mathrm{HF}}, \ket{R_1}, \ket{\Tilde{\nu}}=\ket{\nu}-\ket{R_1}\braket{R_1|\nu} \}$. Here, $\ket{R_1}$ is defined as $\tfrac{1}{\sqrt{2}}\sum_{ai}r_{ai,1}E_{ai}\ket{\mathrm{HF}}$. The Hamiltonian expressed in this full space basis is
\begin{equation}
\begin{aligned}
\mathbf{H}^\mathrm{FS}&=\begin{pmatrix}
\bra{\mathrm{HF}}H\ket{\mathrm{HF}} & \bra{\mathrm{HF}}H\ket{R_1} &  0\\
\bra{R_1}H\ket{\mathrm{HF}} & \bra{\mathrm{R}_1}H\ket{R_1} & \bra{R_1}H\ket{\Tilde{\nu}}\\
0 & \bra{\Tilde{\mu}}H\ket{R_1} & \bra{\Tilde{\mu}}H\ket{\Tilde{\nu}}
\end{pmatrix},
\label{eq:H_fullspace}
\end{aligned}
\end{equation}
where we used that $\bra{\Tilde{\mu}}H\ket{\mathrm{HF}}=\bra{\mathrm{HF}}H\ket{\Tilde{\nu}}=0$. The associated full space generalized eigenvalue problem is defined as
\begin{equation}
    \mathbf{H}^{\mathrm{FS}}\mathbf{x}_n = \mathcal{E}_n \mathbf{S}^{\mathrm{FS}}\mathbf{x}_n
     \label{eq:FSeigen}
\end{equation}
The full space overlap matrix $\mathbf{S}^{\mathrm{FS}}$ is
\begin{equation}
\begin{aligned}
\mathbf{S}^\mathrm{FS}&=\begin{pmatrix}
1 & \braket{\mathrm{HF}|R_1} & \braket{\mathrm{HF}|\Tilde{\nu}} \\
\braket{R_1|\mathrm{HF}} & 1 & \braket{R_1|\Tilde{\nu}}\\
\braket{\Tilde{\mu}|\mathrm{HF}}  & \braket{\Tilde{\mu}|R_1} & \braket{\Tilde{\mu}|\Tilde{\nu}}
\end{pmatrix} =\begin{pmatrix}
1 & 0 & 0\\
\w0 & \w1 & 0\\
0 & 0 & \braket{\Tilde{\mu}|\Tilde{\nu}}
\end{pmatrix}
\end{aligned}
\end{equation}
but it can be replaced with the identity matrix, as shown in the Supplementary Information. The above framework can easily be extended to an arbitrary number of states included in the projector operator. Details can be found in the Supplementary Information.

\clearpage
\subsection*{Applications}

To illustrate the properties of the CVX-HF method, we consider the ammonia molecule. A conical intersection is encountered when one N-H bond is stretched and the molecule is in a nearly planar configuration~\cite{li2007improved,li2014configuration,xu2025conical}. 

\begin{figure}[!htb]
    \centering
    \includegraphics[width=0.97\textwidth]{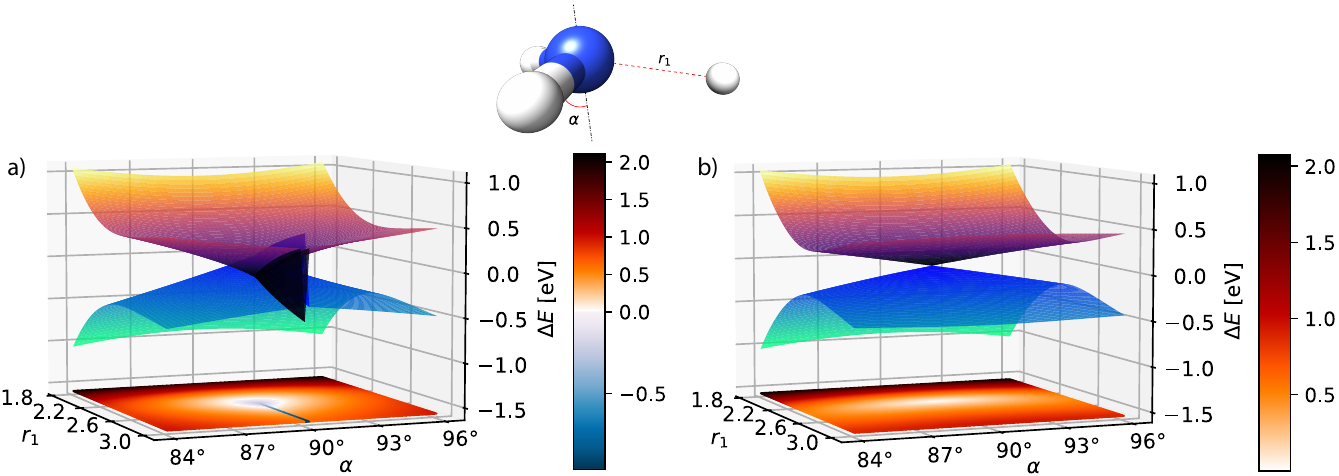}
    \caption{\footnotesize The TDA-TDHF (a) and CVX-HF (b) potential energy surfaces of $\mathrm{S}_0$ and $\mathrm{S}_1$ in ammonia, using aug-cc-pVDZ. For each point the energies are plotted in eV relative to the average energy $\tfrac{1}{2}$($E_0+E_1$) of the states. At the bottom of each plot, a colormap shows the value of $E_1 -E_0$.
    The geometry is shown in the middle, with the bond length $r_1$ and the out-of-plane angle $\alpha$ indicated in red. One N-H bond is stretched while the remaining two are fixed at 1.04 \AA. The out-of-plane angle $\alpha$ is defined as the angle between the vector trisecting the three N-H bonds and any of the N-H bonds. For $\alpha=90^\circ$ the geometry is planar, with all H-N-H angles equal to 120$^\circ$. These angles are kept equal as $\alpha$ changes.}
    \label{fig:NH3_combined_geom}
\end{figure}

In Fig.~\ref{fig:NH3_combined_geom} (a) we show a region close to the intersection for TDA-TDHF, where the HF energy is higher than the excited state energy, resulting in a negative excitation energy shown by a blue region in the colormap at the bottom of the plot. When the geometry is planar ($\alpha = 90^\circ$), the negative excitation energy extends to larger bond lengths, as can be seen from the blue segment in the colormap. 
Looking at the potential energy surfaces, it is clear that the surfaces in this region are not continuous with the surroundings. 
Using CVX-HF (Fig.~\ref{fig:NH3_combined_geom} (b)) projecting the first eigenvector of the Hessian, we obtain continuous energy surfaces, and the intersection is limited to a single point. In the Supplementary Information, we provide a comparison of CVX-HF, TDA-TDHF, GCCSD~\cite{rossi2025generalized}, CCSD and FCI, when stretching one bond at a fixed $\alpha=89.5^\circ$.

\begin{figure}[!htb]
    \centering
    \includegraphics[width=0.97\textwidth]{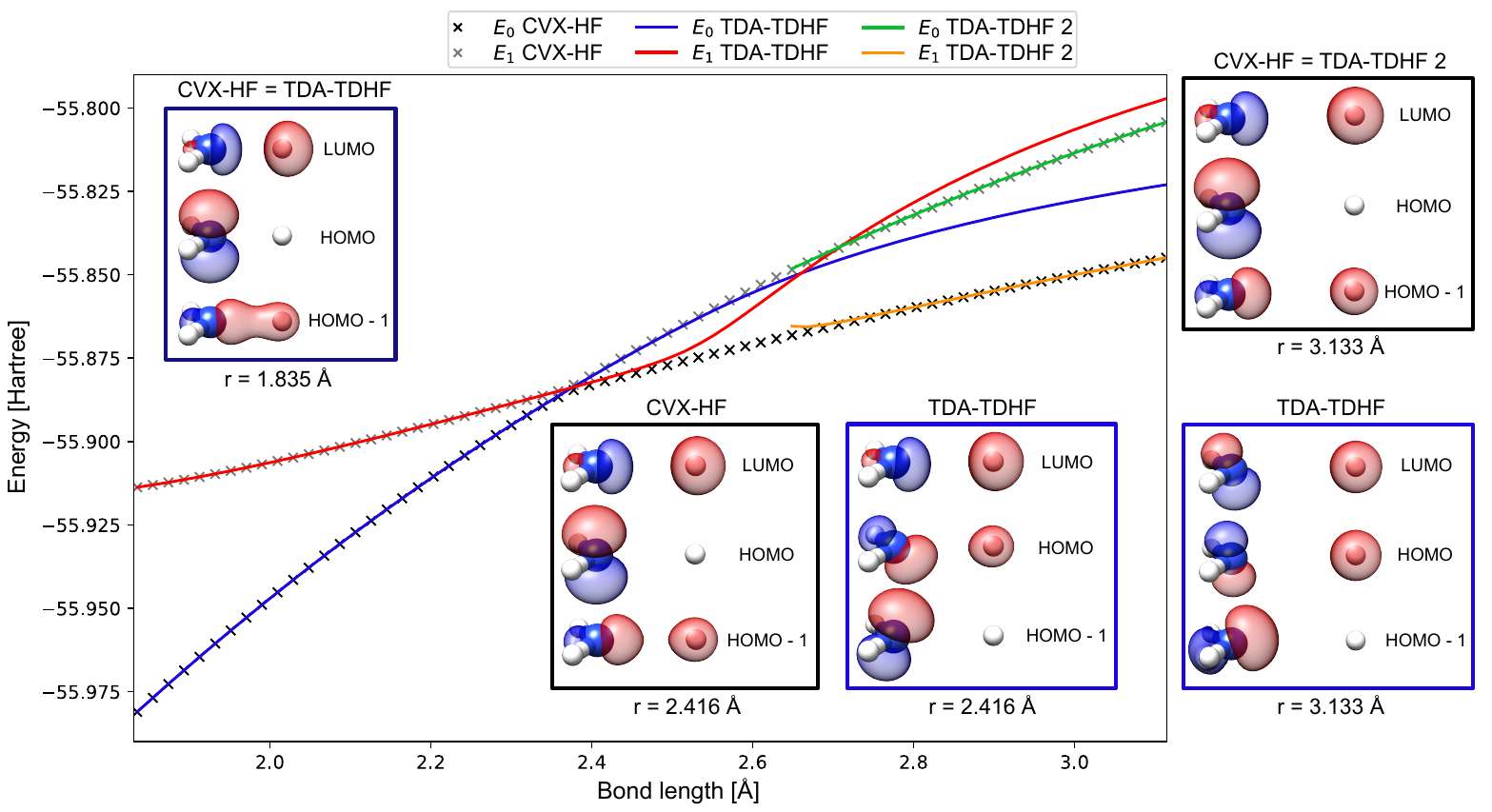}
    \caption{\footnotesize Potential energy curves of states $\mathrm{S}_0$ and $\mathrm{S}_1$ of NH$_3$ computed with the 6-31G* basis set.  One N-H bond is stretched while maintaining a fixed out-of-plane angle of $\alpha = 89.5^\circ$ (see Fig. \ref{fig:NH3_combined_geom} for details). All energies are reported in Hartrees. CVX-HF is shown as black and gray crosses, TDA-TDHF as blue and red lines, and beyond $r = 2.65$ Å, a second Hartree-Fock solution TDA-TDHF 2 is shown with green and orange lines. The HOMO-1, HOMO, and LUMO orbitals are shown for all methods at three representative bond lengths: before ($r = 1.385$ Å), near ($r = 2.416$ Å), and after ($r = 3.133$ Å) the avoided crossing. At $r = 1.385$ Å, only CVX-HF orbitals are shown, as those from TDA-TDHF are visually indistinguishable. Similarly, at $r = 3.133$ Å, only the CVX-HF orbitals are shown, since they closely match those from TDA-TDHF 2.}
    \label{fig:NH3_orbitals}
\end{figure}

It is instructive to analyze the molecular orbitals in the CVX-HF model. 
In Fig.~\ref{fig:NH3_orbitals} we consider the stretching of a single N-H bond at a fixed out-of-plane angle of $89.5^\circ$ and study the changes in energy and character of the $\sigma$ HOMO-1, $p_N$ HOMO and $\sigma^*$ LUMO orbitals. 
For TDA-TDHF we observe two state crossings: at $r = 2.37 $ Å and $r = 2.65$ \AA. Beyond this second point, a new HF solution (TDA-TDHF 2) appears with a negative excitation energy, corresponding to a state that is energetically lower than the original HF solution. In contrast, CVX-HF shows a single avoided crossing at $r = 2.37$\AA. The energy profile shows a smooth and continuous connection from the initial TDA-TDHF solution to the TDA-TDHF 2 after the avoided crossing.
The character of the CVX-HF orbitals is consistent across the entire range of bond lengths. This is not the case for TDA-TDHF, as the $p_N$ orbital gets mixed after the first crossing. The original picture is retrieved with TDA-TDHF 2, only after the second crossing point.

\begin{figure}[!htb]
    \centering
    \includegraphics[width=0.97\textwidth]{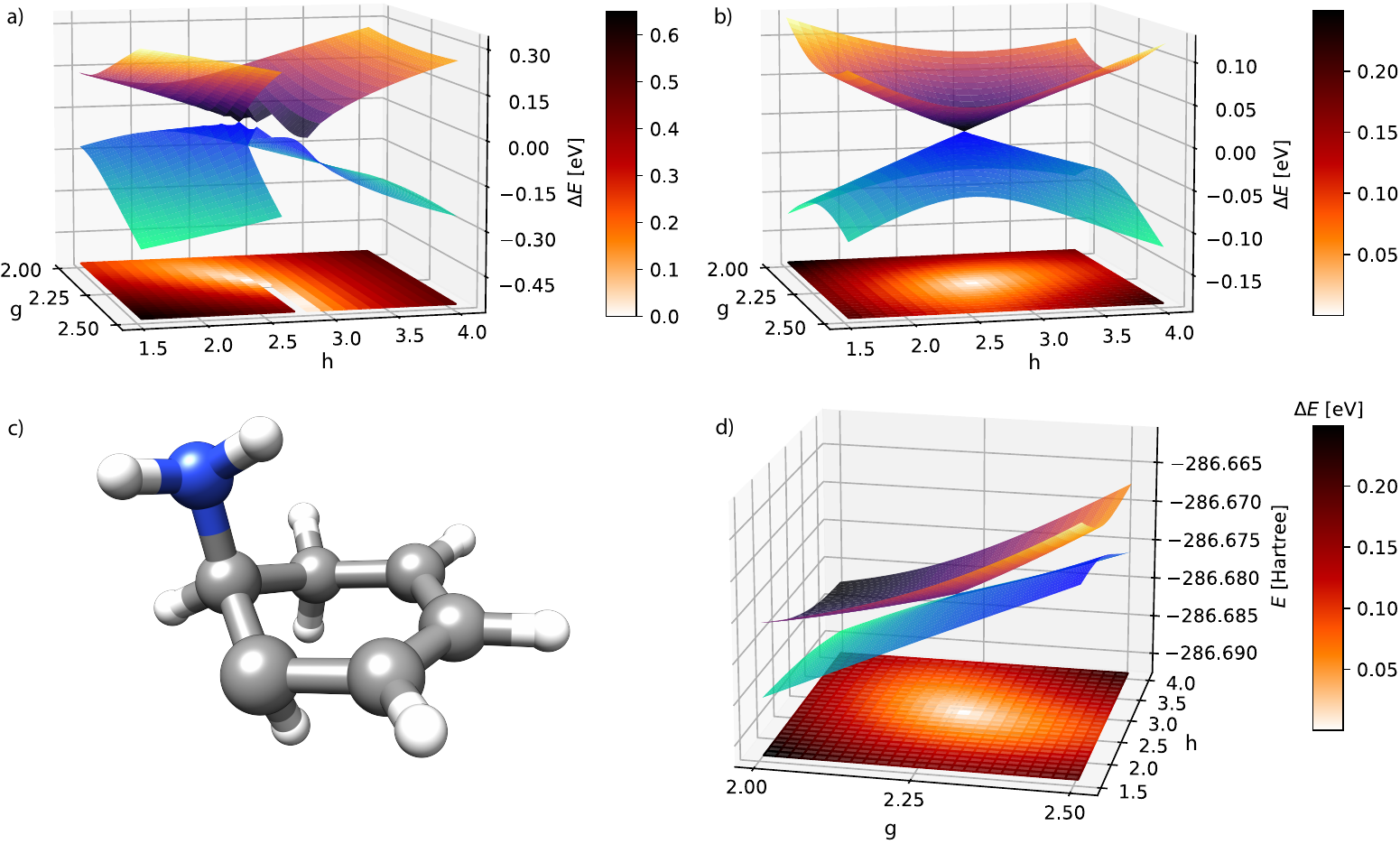}
    \caption{\footnotesize The TDA-TDHF (a) and CVX-HF (b) potential energy curves of $\mathrm{S}_0$ and $\mathrm{S}_1$ in 2,4-cyclohexadien-1-ylamine using cc-pVDZ. For each point the energies are plotted in eV relative to the average energy $\tfrac{1}{2}$($E_0+E_1$) of the states. In (d), the CVX-HF total energy potential energy surfaces for the same system are shown, expressed in Hartree. At the bottom of each plot, a colormap shows the value of $E_1 -E_0$ in eV. The geometry at the CVX-HF CI is shown in (c).}
    \label{fig:1Am_combined}
\end{figure}

We now consider a larger system and study an $S_0$/$S_1$ intersection in 2,4-cyclohexadien-1-ylamine. The $\mathbf{g}$ and $\mathbf{h}$ vectors are obtained with a CCSD algorithm~\cite{kjonstad2023communication} at the geometry in Ref.~\citenum{database_polyene}. Additional information can be found in the Supplementary Information and in Ref.~\citenum{rossi2025generalized}. The potential energy surfaces for TDA-TDHF are shown in Fig.~\ref{fig:1Am_combined} (a). A region in which HF did not converge is present for $g$ greater than 2.25 and $h = 2.75$. The two surfaces are discontinuous across this region, resulting in an overall non-conical shape. In contrast, the results for CVX-HF in Fig.~\ref{fig:1Am_combined} (b) show no convergence issues and the surfaces are continuous, retaining the expected conical shape. 

\begin{figure}[!htb]
    \centering
    \includegraphics[width=0.97\textwidth]{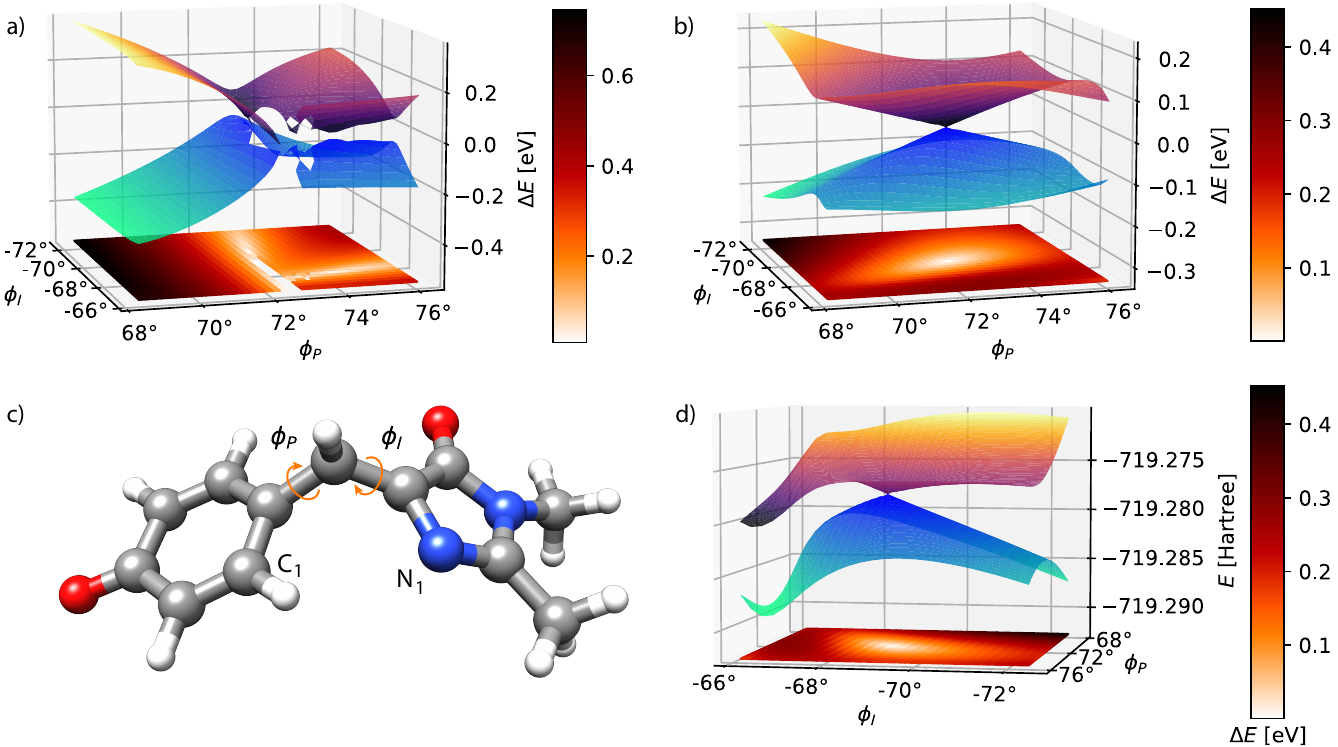}
    \caption{\footnotesize The TDHF (a) and CVX-HF (b) potential energy curves of $\mathrm{S}_0$ and $\mathrm{S}_1$ in GFP using 6-31G*.For each point the energies are plotted in eV relative to the average energy $\tfrac{1}{2}$($E_0+E_1$) of the states. In (d), the CVX-HF total energy potential energy surfaces for the same system are shown, expressed in Hartree. At the bottom of each plot, a colormap shows the value of $E_1 -E_0$ in eV. The geometry at the CVX-HF CI is shown in (c).}
    \label{fig:GFP_combined}
\end{figure}

Lastly, we examine the $S_0$/$S_1$ conical intersection in the anionic form of p-hydroxybenzylidene-2,3-dimethylimidazolinone (HBDI$^-$)~\cite{jones2021resolving}. This molecule is the chromophore of the green fluorescent protein (GFP), a widely used fluorescent marker in biological imaging. Its bright green emission arises from a photoinduced process closely linked to the electronic structure of HBDI$^-$~\cite{tsien1998green,list2024chemical}. The torsional angle $\phi_P$ is defined as the dihedral angle between the carbon atoms of the methine bridge and C$_1$, and $\phi_P$ is defined as the dihedral angle between N$_1$ and the carbon atoms of the methine bridge (see Fig.~\ref{fig:GFP_combined}). The initial structure corresponds to the P90 minimum energy conical intersection (MECI) in Ref.~\citenum{jones2021resolving}, but with modified dihedral angles as described in the Supplementary Information. 
In Fig.~\ref{fig:GFP_combined} (a), a region is observed where HF fails to converge. In addition, the behavior of the potential energy surfaces for $\phi_P > 72.5^\circ$ is significantly different from that at smaller $\phi_P$ values. In the latter case, the energy shows little variation as $\phi_I$ changes. Again, the CVX-HF in Fig. \ref{fig:GFP_combined} (b) retains the correct topology of the intersection.

\clearpage
\section*{Discussion}
In this work, we introduced the Convex Hartree-Fock framework as an efficient method for modeling ground state conical intersections. By reformulating the Hartree-Fock optimization in a subspace where it is convex, the approach yields smooth potential energy surfaces and resolves failures inherent to conventional mean-field methods.
The CVX-HF framework has several important fundamental properties. The method is orbital invariant with respect to rotations within either the occupied or the virtual orbital spaces. Furthermore, as shown in the Supplementary Information, the excitation energies are size-intensive when additional molecules are introduced, such as in solvated systems. The calculated excitation energies are rather insensitive to the number of states that are being projected (see Tables S2 and S3). Therefore, the choice of the number of projected states is not critical and can be adjusted to the number of states involved, for instance, in the dynamics studied.

Despite its advantages, CVX-HF remains fundamentally at the Hartree–Fock level of accuracy, which can limit the reliability for quantitative predictions of excitation energies and electron correlation effects. To overcome this limitation, two strategies can be employed. First,  the orbitals determined in the CVX-HF framework can be used as a basis for more accurate methods, such as coupled cluster models and perturbation theory, ensuring that discontinuities at the HF level do not propagate to the correlated model. This is particularly important close to ground state conical intersections, where coupled cluster displays divergent behavior due to the inability to describe the geometric phase effect~\cite{kjonstad2024understanding}. The CVX-HF orbitals can also be combined with the generalized coupled cluster framework that was developed to give a correct description of ground state conical intersections~\cite{rossi2025generalized}. Second, the framework presented in this work can be applied to TDDFT, using the same exponential parameterization to define the orbitals in the Kohn-Sham equations. The component along the first Hessian eigenvector, determined during the ground state optimization, is removed from the gradient and the $\kappa$ matrix, and later introduced in the final diagonalization step.

A promising feature of the method is that additional excitations can be included in the final diagonalization procedure. A notable example involves systems with multiple chromophores, as in the case of pigments in photosynthetic light-harvesting complexes~\cite{cogdell2006architecture,segatta2019quantum}. When the molecules are infinitely separated, the eigenvectors of the Hessian of the combined system are intensive and will be localized on the individual molecules. These eigenvectors may be used to construct a size-extensive parameterization.
As an example, consider the case of two non-interacting systems, $A$ and $B$. The wave function of the combined system should match the product of the wave functions of the isolated systems. For CVX-HF, the wave functions of the two isolated systems in the reduced space can be expressed as
\begin{align}
&\ket{\psi_A}= \ket{\text{HF}_A}x_0^A + \ket{R_A}x_1^A \\
&\ket{\psi_B}= \ket{\text{HF}_B}x_0^B + \ket{R_B}x_1^B .
\end{align}
The wave function for the combined system in the reduced space, when projecting 2 states, is given by
\begin{equation}
\ket{\psi}= \ket{\text{HF}}x_0 + \ket{R_1}x_1  + \ket{R_2}x_2 \,,
\end{equation}
where
\begin{align}
&\ket{\text{HF}}=\ket{\text{HF}_A}\ket{\text{HF}_B}\\
&\ket{R_1}=\ket{R_A}\ket{\text{HF}_B}\\
&\ket{R_2}=\ket{\text{HF}_A}\ket{R_B} \,.
\end{align}

This wave function is not size-extensive because the component given by the combined excitation $\ket{R_1 R_2}=\ket{R_A}\ket{R_B}$ is missing. In order to retrieve the correct product form, this term can be included in the final reduced space eigenvalue problem, only changing the prefactor of the overall CVX-HF scaling. The size-extensivity of the combined system in the extended space is shown in the Supplementary Information. The same idea can be generalized to a larger number of chromophores. Finally, we note that, in some cases, double excitations are needed in order to describe the conical intersection. These can be added in a similar manner.

\section*{Methods}\label{sec:methods}
All calculations reported in this work were performed using an implementation of the CVX-HF framework in a local development version of the eT program~\cite{folkestad20201}. The algorithm presented below outlines the iterative procedure employed in CVX-HF to self-consistently solve the Hessian eigenvalue problem and optimize the ground state within the restricted subspace. The full space eigenvalue problem is solved once at the end of the procedure. Additional details on the definitions of vectors and matrices are provided in the Supplementary Information.

\begin{algorithm}[!htb]
\begin{algorithmic}[1]
\State $C_0 \gets$ SAD guess
\State Initialize $n = 0$, $ \bm\kappa^{[0]} = \mathbf{0}$ and $ C^{[0]}=C_0$
\While{($n<n_\text{max}$ \textbf{and} $\lVert\tilde{\mathbf{G}}^{(0)}\rVert_{L_2} > $ threshold)}
    \State Solve the eigenvalue problem $\mathbf{G}^{(1)[n]}\mathbf{r}_i^{[n]} = \lambda_i^{[n]} \mathbf{r}_i^{[n]}$, \quad $i=1, \dots, n_\text{excited}$
    \State Construct $\mathbf{G}^{(0)[n]}$ and $\hat{P}^{[n]}=\mathcal{I} -\sum_{i=1}^{n_\text{proj}} \mathbf{r}_i^{[n]} \mathbf{r}_i^{[n]T}$
    \State $\tilde{\mathbf{G}}^{(0)[n]} \gets \hat{P}^{[n]} \mathbf{G}^{(0)[n]}$
    \State Solve $\hat{P}^{[n]}\mathbf{G}^{(1)[n]} \hat{P}^{[n]}\Delta \bm\kappa^{[n]} = - \tilde{\mathbf{G}}^{(0)[n]}$ for $\Delta \bm\kappa^{[n]}$
    \State $\bm\kappa^{[n+1]} \gets \hat{P}^{[n]} (\bm\kappa^{[n]} + \Delta \bm\kappa^{[n]})$ 
    \State $C^{[n+1]}\gets C_0  \,\,\text{exp}(\bm{\kappa}^{[n+1]})$
    \State $ n  \gets n+1$
\EndWhile
\State Construct and diagonalize the reduced matrix $\mathbf{H}^\mathrm{RS}$
\State Solve the final eigenproblem $\mathbf{H}^\mathrm{FS} \mathbf{x}_j = \mathcal{E}_j \mathbf{x}_j$,\quad $j = 1, \dots, n_\text{excited}+1$
\end{algorithmic}
\caption{CVX-HF algorithm}\label{alg:chf}
\end{algorithm}
The eigenvalue problem in step 4 is solved using Davidson's algorithm, avoiding the explicit calculation of second-derivative terms. In each iteration, the threshold is updated as $\text{min}(10^{-2}, \lVert\tilde{\mathbf{G}}^{(0)[n]}\rVert_{L_2})$ to match the convergence level of the ground state equations. This approach ensures that the computational cost of this step is distributed across the entire cycle, instead of unnecessarily solving the problem to full accuracy at every iteration.
Step 7 is solved using a trust-region algorithm, again avoiding the explicit calculation of second-derivative terms. The final eigenvalue problem in step 13 is also solved using Davidson's algorithm, initialized with the eigenvectors obtained from the direct diagonalization in step 12. When multiple calculations are performed for the same system at similar geometries, the final value of $\bm\kappa$ from a previous calculation can be used as the initial guess for the next calculation. This requires consistency among the SAD guesses of both calculations, which is ensured through a diabatization step of the new $C_0$ coefficients with respect to those of the previous calculation.

\section*{Acknowledgments}
We thank Eirik F. Kjønstad for enlightening discussions. This work was supported by the European Research Council (ERC) under the European Union's Horizon 2020 Research and Innovation Program (grant agreement No. 101020016).

\section*{Author contributions}
F.R. and H.K. conceived the CVX-HF framework, analyzed the data and wrote the paper. F.R. developed the implementation in eT and performed all the calculations. H.K. supervised the project.

\section*{Competing interests}
The authors declare no competing interests.

\bibliography{sn-bibliography}



\section*{Supplementary material}

# Supplementary Information for "Convex Hartree-Fock theory: A simple framework for ground state conical intersections"

Federico Rossi[1] and Henrik Koch*[1]*

[1]Department of Chemistry, Norwegian University of Science and Technology, Trondheim, 7491, Norway.

*Corresponding author(s). E-mail(s): henrik.koch@ntnu.no;

# Contents

# S1 Hamiltonian matrix in the new basis

## S1.1 Single projected state

The full space Hamiltonian matrix with a single projected state is

$$\mathbf{H}^{\text{FS}} = \begin{pmatrix} \langle \text{HF}|H|\text{HF}\rangle & \langle \text{HF}|H|\text{R}_1\rangle & \langle \text{HF}|H|\tilde{\nu}\rangle \\ \langle \text{R}_1|H|\text{HF}\rangle & \langle \text{R}_1|H|\text{R}_1\rangle & \langle \text{R}_1|H|\tilde{\nu}\rangle \\ \langle \tilde{\mu}|H|\text{HF}\rangle & \langle \tilde{\mu}|H|\text{R}_1\rangle & \langle \tilde{\mu}|H|\tilde{\nu}\rangle \end{pmatrix} = \begin{pmatrix} E_0 & \mathbf{F}_{OV}\mathbf{r}_1 & 0 \\ \mathbf{F}_{OV}\mathbf{r}_1 & E_0 + \mathbf{r}_1^T\mathbf{A}\mathbf{r}_1 & V_\nu \\ 0 & V_\mu & Z_{\mu\nu} + \delta_{\mu\nu}E_0 \end{pmatrix} \quad (1)$$

where

$$V_\mu = \langle \tilde{\mu}|H|\text{R}_1\rangle = \langle \mu|H|\text{R}_1\rangle - \langle \mu|\text{R}_1\rangle\langle \text{R}_1|H|\text{R}_1\rangle \quad (2)$$

$$= \sum_\nu (E_0\delta_{\mu\nu} + A_{\mu\nu})r_{1,\nu} - r_{1,\mu}(E_0 + \mathbf{r}_1^T\mathbf{A}\mathbf{r}_1) \quad (3)$$

$$= A_{\mu\nu}r_{1,\nu} - r_{1,\mu}\mathbf{r}_1^T\mathbf{A}\mathbf{r}_1 \quad (4)$$

$$Z_{\mu\nu} = \langle \tilde{\mu}|H|\tilde{\nu}\rangle - \delta_{\mu\nu}E_0 = \langle \mu|H|\nu\rangle - \delta_{\mu\nu}E_0 - \langle \tilde{\mu}|H|\text{R}_1\rangle\langle \text{R}_1|\nu\rangle + \quad (5)$$

$$- \langle \mu|\text{R}_1\rangle\langle \text{R}_1|H|\tilde{\mu}\rangle - \langle \mu|\text{R}_1\rangle\langle \text{R}_1|H|\text{R}_1\rangle\langle \text{R}_1|\nu\rangle \quad (6)$$

$$= A_{\mu\nu} - r_{1,\mu}V_\nu - V_\mu r_{1,\nu} - r_{1,\mu}r_{1,\nu}(E_0 + \mathbf{r}_1^T\mathbf{A}\mathbf{r}_1) \quad (7)$$

## S1.2 Multiple projected states

The reduced space Hamiltonian matrix with $N$ projected states is defined as

$$\mathbf{H}^{\text{RS}} = \begin{pmatrix} \langle \text{HF}|H|\text{HF}\rangle & \langle \text{HF}|H|\text{R}_1\rangle & \ldots & \langle \text{HF}|H|\text{R}_N\rangle \\ \langle \text{R}_1|H|\text{HF}\rangle & \langle \text{R}_1|H|\text{R}_1\rangle & \ldots & \langle \text{R}_1|H|\text{R}_N\rangle \\ \vdots & \vdots & \ddots & \vdots \\ \langle \text{R}_N|H|\text{HF}\rangle & \langle \text{R}_N|H|\text{R}_1\rangle & \ldots & \langle \text{R}_N|H|\text{R}_N\rangle \end{pmatrix} = \begin{pmatrix} E_0 & \mathbf{F}_{OV}\mathbf{r}_1 & \ldots & \mathbf{F}_{OV}\mathbf{r}_N \\ \mathbf{F}_{OV}\mathbf{r}_1 & E_0 + \mathbf{r}_1^T\mathbf{A}\mathbf{r}_1 & \ldots & \mathbf{r}_1^T\mathbf{A}\mathbf{r}_N \\ \vdots & \vdots & \ddots & \vdots \\ \mathbf{F}_{OV}\mathbf{r}_N & \mathbf{r}_N^T\mathbf{A}\mathbf{r}_1 & \ldots & E_0 + \mathbf{r}_N^T\mathbf{A}\mathbf{r}_N \end{pmatrix} \quad (8)$$

where

$$W_{IJ} = \langle \text{R}_I|H|\text{R}_J\rangle - \delta_{IJ}E_0 = \mathbf{r}_I^T\mathbf{A}\mathbf{r}_J \quad (9)$$

The full space Hamiltonian matrix reads

$$\mathbf{H}^{\text{FS}} = \begin{pmatrix} \langle \text{HF}|H|\text{HF}\rangle & \langle \text{HF}|H|\text{R}_1\rangle & \ldots & \langle \text{HF}|H|\text{R}_N\rangle & \langle \text{HF}|H|\tilde{\nu}\rangle \\ \langle \text{R}_1|H|\text{HF}\rangle & \langle \text{R}_1|H|\text{R}_1\rangle & \ldots & \langle \text{R}_1|H|\text{R}_N\rangle & \langle \text{R}_1|H|\tilde{\nu}\rangle \\ \vdots & \vdots & \ddots & \vdots & \vdots \\ \langle \text{R}_N|H|\text{HF}\rangle & \langle \text{R}_N|H|\text{R}_1\rangle & \ldots & \langle \text{R}_N|H|\text{R}_N\rangle & \langle \text{R}_N|H|\tilde{\nu}\rangle \\ \langle \tilde{\mu}|H|\text{HF}\rangle & \langle \tilde{\mu}|H|\text{R}_1\rangle & \ldots & \langle \tilde{\mu}|H|\text{R}_N\rangle & \langle \tilde{\mu}|H|\tilde{\nu}\rangle \end{pmatrix} = \quad (10)$$

$$
= \begin{pmatrix}
E_0 & \mathbf{F}_{OV}\mathbf{r}_1 & \ldots & \mathbf{F}_{OV}\mathbf{r}_N & X_\nu \\
\mathbf{F}_{OV}\mathbf{r}_1 & E_0 + \mathbf{r}_1^T \mathbf{A}\mathbf{r}_1 & \ldots & \mathbf{r}_1^T \mathbf{A}\mathbf{r}_N & Y_{1,\nu} \\
\vdots & \vdots & \ddots & \vdots & \vdots \\
\mathbf{F}_{OV}\mathbf{r}_N & \mathbf{r}_N^T \mathbf{A}\mathbf{r}_1 & \ldots & E_0 + \mathbf{r}_N^T \mathbf{A}\mathbf{r}_N & Y_{N,\nu} \\
0 & V_{1,\mu} & \ldots & V_{N,\mu} & Z_{\mu\nu} + \delta_{\mu\nu} E_0
\end{pmatrix} \quad (11)
$$

where

$$
X_\nu = \langle \mathrm{HF} | H | \tilde{\nu} \rangle = \eta_\nu - \sum_I (\boldsymbol{\eta}^T \mathbf{r}_I) \mathrm{R}_{I\nu} \quad (12)
$$

$$
Y_{I,\nu} = \langle \mathrm{R}_I | H | \tilde{\nu} \rangle = \langle \mathrm{R}_I | H | \nu \rangle - \sum_J (E_0 \delta_{IJ} + W_{IJ}) \mathrm{R}_{J\nu} \quad (13)
$$

$$
V_{I,\mu} = \langle \tilde{\mu} | H | \mathrm{R}_I \rangle = \langle \mu | H | \mathrm{R}_I \rangle - \sum_J \mathrm{R}_{J\mu} (E_0 \delta_{IJ} + W_{JI}) \quad (14)
$$

$$
Z_{\mu\nu} = \langle \tilde{\mu} | H | \tilde{\nu} \rangle = \langle \mu | H | \nu \rangle - \delta_{\mu\nu} E_0 - \sum_I \mathrm{R}_{I\mu} Y_{I,\nu} - \sum_I V_{I,\mu} \mathrm{R}_{I\nu} - \sum_{IJ} \mathrm{R}_{I\mu}(E_0 \delta_{IJ} + W_{IJ}) \mathrm{R}_{J\nu} \quad (15)
$$

## S2 Metric matrix

Taking as example the case of a single projected state, we show that the inclusion of the metric matrix is not needed when solving the full space eigenvalue problem

$$\mathbf{H}^{\mathrm{FS}}\mathbf{x}_n = \mathcal{E}_n \mathbf{S}^{\mathrm{FS}}\mathbf{x}_n. \tag{16}$$

We report the matrices defined as

$$\mathbf{H}^{\mathrm{FS}} = \begin{pmatrix} \langle \mathrm{HF}|H|\mathrm{HF}\rangle & \langle \mathrm{HF}|H|\mathrm{R}_1\rangle & 0 \\ \langle \mathrm{R}_1|H|\mathrm{HF}\rangle & \langle \mathrm{R}_1|H|\mathrm{R}_1\rangle & \langle \mathrm{R}_1|H|\tilde{\nu}\rangle \\ 0 & \langle \tilde{\mu}|H|\mathrm{R}_1\rangle & \langle \tilde{\mu}|H|\tilde{\nu}\rangle \end{pmatrix} \tag{17}$$

$$\mathbf{S}^{\mathrm{FS}} = \begin{pmatrix} 1 & \langle \mathrm{HF}|R_1\rangle & \langle \mathrm{HF}|\tilde{\nu}\rangle \\ \langle R_1|\mathrm{HF}\rangle & 1 & \langle R_1|\tilde{\nu}\rangle \\ \langle \tilde{\mu}|\mathrm{HF}\rangle & \langle \tilde{\mu}|R_1\rangle & \langle \tilde{\mu}|\tilde{\nu}\rangle \end{pmatrix} = \begin{pmatrix} 1 & 0 & 0 \\ 0 & 1 & 0 \\ 0 & 0 & \langle \tilde{\mu}|\tilde{\nu}\rangle \end{pmatrix}. \tag{18}$$

In what follows, we will show that the eigenvectors of the standard (non-generalized) eigenvalue problem

$$\mathbf{H}^{\mathrm{FS}}\mathbf{x}_n = \mathcal{E}_n \mathbf{x}_n \tag{19}$$

are such that $\mathbf{S}^{\mathrm{FS}}\mathbf{x}_n = \mathbf{x}_n$, which means that they are also eigenvectors of the generalized eigenvalue problem with the same eigenvalue.

We start by writing explicitly $\mathbf{S}^{\mathrm{FS}}\mathbf{x} - \mathbf{x}$:

$$\mathbf{S}^{\mathrm{FS}}\mathbf{x} - \mathbf{x} = \begin{pmatrix} 1 & 0 & 0 \\ 0 & 1 & 0 \\ 0 & 0 & \langle \tilde{\mu}|\tilde{\nu}\rangle \end{pmatrix} \begin{pmatrix} x_0 \\ x_1 \\ x_\nu \end{pmatrix} - \begin{pmatrix} x_0 \\ x_1 \\ x_\mu \end{pmatrix} = \begin{pmatrix} 0 \\ 0 \\ -r_{1,\mu}r_{1,\nu}x_\nu \end{pmatrix} \tag{20}$$

We are now left to show that $\sum_\mu r_{1,\mu} x_\mu = 0$. Since $\mathbf{x}$ is an eigenvector of $\mathbf{H}^{\mathrm{FS}}$, we know that

$$\langle \tilde{\mu}|\bar{H}|R_1\rangle x_1 + \sum_\nu \langle \tilde{\mu}|\bar{H}|\tilde{\nu}\rangle x_\nu = \mathcal{E} x_\mu \tag{21}$$

and we can isolate $x_\mu$ and substitute

$$\sum_\mu r_{1,\mu} x_\mu = \frac{1}{\mathcal{E}} \sum_\mu r_{1,\mu} \langle \tilde{\mu}| \left( \bar{H}|R_1\rangle x_1 + \sum_\nu \bar{H}|\tilde{\nu}\rangle x_\nu \right) \tag{22}$$

$$= \frac{1}{\mathcal{E}} \sum_\mu r_{1,\mu} (\langle \mu| - r_{1,\mu}\langle R_1|) \left( \bar{H}|R_1\rangle x_1 + \sum_\nu \bar{H}|\tilde{\nu}\rangle x_\nu \right) \tag{23}$$

$$= \frac{1}{\mathcal{E}} (\langle R_1| - \langle R_1|) \left( \bar{H}|R_1\rangle x_1 + \sum_\nu \bar{H}|\tilde{\nu}\rangle x_\nu \right) = 0. \tag{24}$$

Note that we assumed $\mathcal{E} \neq 0$ and $\sum_\mu r_{1,\mu} r_{1,\mu} = 1$.

## S3 Reduced matrix with additional vectors

We consider the case of two molecules $A$ and $B$ infinitely separated. The CVX-HF reduced space matrices for the two separated systems are defined as

$$\mathbf{H}_A^{\text{RS}} = \begin{pmatrix} \langle \text{HF}_A | H_A | \text{HF}_A \rangle & \langle \text{HF}_A | H_A | R_A \rangle \\ \langle R_A | H_A | \text{HF}_A \rangle & \langle R_A | H_A | R_A \rangle \end{pmatrix} = \begin{pmatrix} E_0^A & c_A \\ c_A & E_1^A \end{pmatrix} \quad (25)$$

$$\mathbf{H}_B^{\text{RS}} = \begin{pmatrix} \langle \text{HF}_B | H_B | \text{HF}_B \rangle & \langle \text{HF}_B | H_B | R_B \rangle \\ \langle R_B | H_B | \text{HF}_B \rangle & \langle R_B | H_B | R_B \rangle \end{pmatrix} = \begin{pmatrix} E_0^B & c_B \\ c_B & E_1^B \end{pmatrix}. \quad (26)$$

For the total system, the reduced matrix extended to include the combined excitation reads

$$\mathbf{H}^{\text{RS}} = \begin{pmatrix} \langle \text{HF} | H | \text{HF} \rangle & \langle \text{HF} | H | R_1 \rangle & \langle \text{HF} | H | R_2 \rangle & \langle \text{HF} | H | R_1 R_2 \rangle \\ \langle R_1 | H | \text{HF} \rangle & \langle R_1 | H | R_1 \rangle & \langle R_1 | H | R_2 \rangle & \langle R_1 | H | R_1 R_2 \rangle \\ \langle R_2 | H | \text{HF} \rangle & \langle R_2 | H | R_1 \rangle & \langle R_2 | H | R_2 \rangle & \langle R_2 | H | R_1 R_2 \rangle \\ \langle R_1 R_2 | H | \text{HF} \rangle & \langle R_1 R_2 | H | R_1 \rangle & \langle R_1 R_2 | H | R_2 \rangle & \langle R_1 R_2 | H | R_1 R_2 \rangle \end{pmatrix}. \quad (27)$$

Assuming the ordering of the eigenvectors is such that $|R_1\rangle$ is the one located on system A, at infinite distance we have

$$|\text{HF}\rangle = |\text{HF}_A\rangle |\text{HF}_B\rangle \quad (28)$$
$$|R_1\rangle = |R_A\rangle |\text{HF}_B\rangle \quad (29)$$
$$|R_2\rangle = |\text{HF}_A\rangle |R_B\rangle \quad (30)$$
$$|R_1 R_2\rangle = |R_A\rangle |R_B\rangle. \quad (31)$$

Using this, the extended reduced matrix for the combined system takes the form

$$\mathbf{H}^{\text{RS}} = \begin{pmatrix} E_0^A + E_0^B & c_A & c_B & 0 \\ c_A & E_1^A + E_0^B & 0 & c_B \\ c_B & 0 & E_0^A + E_1^B & c_A \\ 0 & c_B & c_A & E_1^A + E_1^B \end{pmatrix}. \quad (32)$$

This shows that $\mathbf{H}^{\text{RS}}$ can be expressed as the Kronecker sum

$$\mathbf{H}_B^{\text{RS}} \oplus \mathbf{H}_A^{\text{RS}} = \mathbf{H}_B^{\text{RS}} \otimes \mathbf{I}_2 + \mathbf{I}_2 \otimes \mathbf{H}_A^{\text{RS}}. \quad (33)$$

The eigenvalues of such a Kronecker sum are obtained as the pairwise sum of the eigenvalues of $\mathbf{H}_B^{\text{RS}}$ and $\mathbf{H}_A^{\text{RS}}$. (Theorem 10.1 [1])

# S4 Extensivity

The extensivity of the method is tested numerically on a single molecule of 2,4-cyclohexadien-1-ylamine, adding He atoms at large distances.

| $n_{He}$ | $E_0$ | $E_1$ | $E_0 - n_{He}E_{He}$ | $E_1 - n_{He}E_{He}$ |
|---|---|---|---|---|
| 0 | -286.71831598 | -286.64708752 | -286.71831598 | -286.64708752 |
| 1 | -289.57347646 | -289.50224800 | -286.71831598 | -286.64708752 |
| 2 | -292.42863694 | -292.35740848 | -286.71831598 | -286.64708752 |
| 3 | -295.28379741 | -295.21256895 | -286.71831598 | -286.64708752 |

**Table S1**: CVX-HF energy in Hartrees of ground and first excited states of 2,4-cyclohexadien-1-ylamine, adding multiple He atoms separated by 500 a.u. in all directions. On the right, the same energies are reported after removing the HF energy for the corresponding number of isolated He atoms. For a single He atom, $E_{He} = -2.855160477$ Hartree. The geometry is reported in Table S5. The basis is cc-pVDZ and the converging thresholds are set at $10^{-8}$ with a single projected state.

# S5 Effect of multiple projections

To study the effect of the inclusion of multiple Hessian eigenvectors in the projection operator, the CVX-HF ground and excited state energies are calculated after progressively increasing the number of projected states $n_{proj}$. The results for the full matrix energies and excitation energies are reported below.

| $n_{proj}$ | $E_0$ | $E_1$ | $E_2$ | $E_3$ | $E_4$ | $E_5$ |
|---|---|---|---|---|---|---|
| 1 | -719.277870 | -719.277718 | - | - | - | - |
| 2 | -719.277725 | -719.277717 | -719.236711 | - | - | - |
| 3 | -719.277313 | -719.276764 | -719.236363 | -719.128231 | - | - |
| 4 | -719.275613 | -719.274790 | -719.234622 | -719.127736 | -719.114103 | - |
| 5 | -719.275568 | -719.274900 | -719.234536 | -719.128065 | -719.114284 | -719.108815 |

**Table S2**: CVX-HF energy in Hartrees of ground and excited states of HBDI$^-$ when increasing the number of vectors included in the projection operator. The geometry is reported in Table S9. The basis is 6-31G* and the converging thresholds are set at $10^{-6}$.

| $n_{proj}$ | $\omega_1$ | $\omega_2$ | $\omega_3$ | $\omega_4$ | $\omega_5$ |
|---|---|---|---|---|---|
| 1 | 0.004151 | - | - | - | - |
| 2 | 0.000212 | 1.116030 | - | - | - |
| 3 | 0.014954 | 1.114308 | 4.056745 | - | - |
| 4 | 0.022385 | 1.115429 | 4.023947 | 4.394918 | - |
| 5 | 0.018170 | 1.116529 | 4.013757 | 4.388774 | 4.537569 |

**Table S3**: CVX-HF excitation energies in eV for HBDI$^-$ when increasing the number of vectors included in the projection operator. The geometry is reported in Table S9. The basis is 6-31G* and the converging thresholds are set at $10^{-6}$.

# S6 Geometries

All 2D scans are run with the initial geometry, **g** and **h** vectors reported in the following. These vectors are originally determined in Hartree/Bohr and later used as displacement vectors in Bohr, defining a new geometry $\mathbf{r}_0 + \alpha \mathbf{g} + \beta \mathbf{h}$ from the initial geometry $\mathbf{r}_0$.

Conversion factors: 1Å = 1.8897259886 Bohr, 1 Hartree = 27.2114079527 eV

## S6.1 Ammonia

Table S4: Ammonia CVX-HF/aug-cc-pVDZ $S_0/S_1$ conical intersection geometry in Angstrom, corresponding to $r_1 = 2.375131$ Å and $\alpha = 90°$.

| Atom | $x$ | $y$ | $z$ |
|---|---|---|---|
| N | 0.00000000 | 0.00000000 | 0.00000000 |
| H | 2.37513100 | 0.00000000 | -0.00000000 |
| H | -0.52000000 | 0.90066642 | -0.00000000 |
| H | -0.52000000 | -0.90066642 | -0.00000000 |

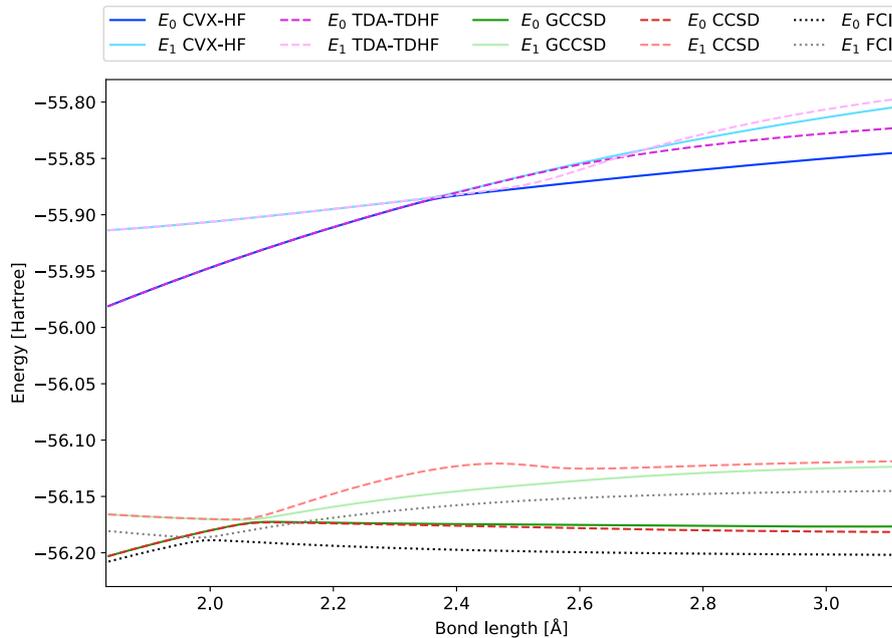

Fig. S1: Potential energy curves of $S_0$ and $S_1$ in $NH_3$ using 6-31G* for different methods. One N-H bond length is stretched with a constant out-of-plane angle $\alpha = 89.5°$ (see Fig. 1 for more information). All energies are expressed in Hartrees.

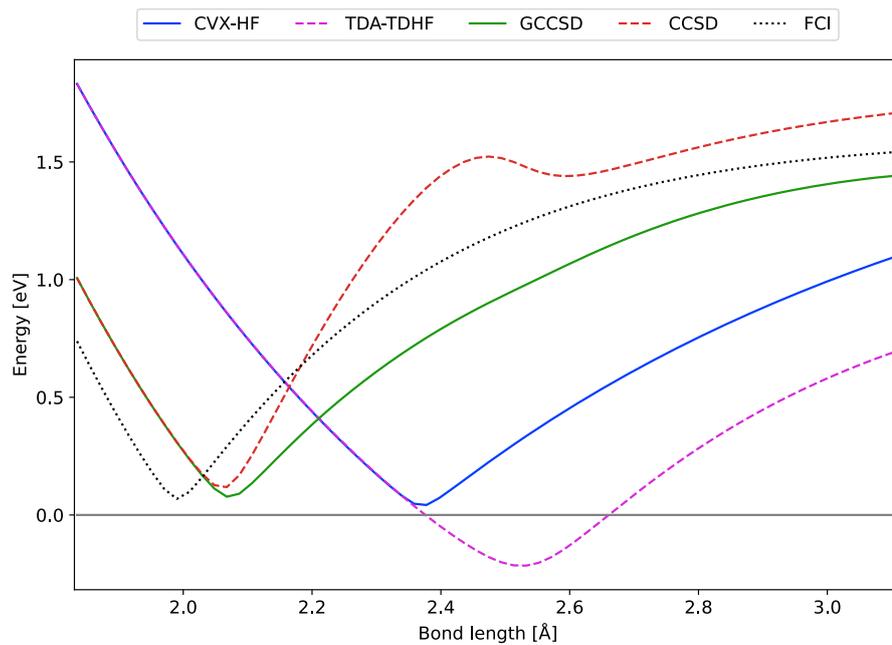

**Fig. S2**: Excitation energy for the $S_0$-$S_1$ transition in $NH_3$ using 6-31G* for different methods. One N-H bond length is stretched with a constant out-of-plane angle $\alpha = 89.5°$ (see Fig. 1 for more information). All energies are expressed in eV.

## S6.2 2,4-cyclohexadien-1-ylamine

The initial structure is obtained from Ref. 2 from which the **g** and **h** vectors are calculated at CCSD/6-31G level [3]. Additional details can be found in Ref. 4.

Table **S5**: 2,4-Cyclohexadien-1-ylamine initial geometry in Bohr.

| Atom | $x$ | $y$ | $z$ |
| --- | --- | --- | --- |
| N | 2.485229105603 | 0.547832318017 | -1.016626362773 |
| C | -1.020180309090 | -2.514973398583 | -1.130137584081 |
| C | 0.000000000000 | 0.000000000000 | 0.000000000000 |
| C | -0.152827830887 | -4.842609039125 | 0.243271218683 |
| C | 0.000000000000 | 0.000000000000 | 2.948948714717 |
| C | 0.393673950618 | -4.812906418385 | 2.772429176890 |
| C | 0.638744772120 | -2.397590397164 | 4.027981665337 |
| H | 3.068814728580 | 2.219087393160 | -0.340105507921 |
| H | 3.743103862610 | -0.735412398662 | -0.407797247778 |
| H | -0.597826785001 | -2.656733416686 | -3.131827502962 |
| H | -1.241628965952 | 1.485927944692 | -0.652347722798 |
| H | -3.073407175968 | -2.431583674518 | -0.951269929831 |
| H | -0.051896980885 | -6.598968101875 | -0.785039323552 |
| H | -2.020803718174 | -1.042080601584 | 3.045064771778 |
| H | 0.977265443085 | -6.512166056261 | 3.733568986966 |
| H | 1.430957011893 | -2.465032088650 | 5.918242701070 |

Table **S6**: 2,4-Cyclohexadien-1-ylamine **g** vector in Hartree/Bohr.

| Atom | $x$ | $y$ | $z$ |
| --- | --- | --- | --- |
| N | 0.004971062452 | -0.000006312063 | -0.008307843382 |
| C | -0.016216875814 | -0.011028497968 | 0.013811359333 |
| C | 0.028843656521 | 0.052451171540 | 0.008092929065 |
| C | -0.002450100320 | -0.007281793260 | -0.010822870907 |
| C | -0.081339063409 | -0.074869115270 | -0.049735972788 |
| C | -0.013448003251 | 0.032461991297 | 0.014876665117 |
| C | 0.068149686093 | -0.043067069541 | 0.067549127712 |
| H | 0.000405335195 | -0.001041323709 | -0.002325561626 |
| H | -0.001324788614 | -0.000710206799 | -0.000031584236 |
| H | 0.003448251125 | 0.006602895615 | 0.002116067133 |
| H | -0.001001097065 | -0.000044020470 | -0.002328844295 |
| H | 0.003064563939 | 0.003521849557 | -0.003645421539 |
| H | 0.001814047671 | 0.001543968627 | -0.001141347406 |
| H | -0.009685201838 | 0.038125491323 | -0.021492693576 |
| H | -0.000178824202 | 0.003941613764 | 0.002797133980 |
| H | 0.014947351518 | -0.000600642643 | -0.009411142586 |

Table S7: 2,4-Cyclohexadien-1-ylamine **h** vector in Hartree/Bohr.

| Atom | $x$ | $y$ | $z$ |
|------|-----|-----|-----|
| N | 0.002487591355 | 0.001059423611 | -0.001229774795 |
| C | 0.001417281298 | 0.007961593644 | 0.001946831613 |
| C | 0.002788554845 | -0.012082019030 | -0.001934955082 |
| C | 0.000037767487 | -0.003593455041 | 0.002688939505 |
| C | -0.010315266883 | -0.006727478406 | -0.008749189841 |
| C | 0.018041726684 | 0.000359400262 | -0.009740084068 |
| C | 0.007161006545 | -0.003086550689 | -0.006418446171 |
| H | 0.000523768377 | 0.000163090505 | 0.000341198029 |
| H | 0.000201834734 | 0.000043417177 | -0.000047450886 |
| H | 0.000287627165 | 0.001485980656 | 0.000635812663 |
| H | 0.000112190184 | 0.000739350784 | 0.000996160267 |
| H | -0.001339158720 | -0.002603931746 | -0.001285345959 |
| H | -0.003490679658 | -0.000662035975 | 0.000729035443 |
| H | -0.004741496715 | 0.016752015594 | 0.017562156627 |
| H | -0.001039863972 | -0.000167499259 | 0.000074916208 |
| H | -0.012149841791 | 0.000529111955 | 0.004488870984 |

Table S8: 2,4-Cyclohexadien-1-ylamine $S_0/S_1$ CI for CVX-HF/cc-pvdz in Bohr. The geometry corresponds to $(g, h) = (2.2662, 2.7257)$.

| Atom | $x$ | $y$ | $z$ |
|------|-----|-----|-----|
| N | 2.503274956386 | 0.550705684763 | -1.038805596200 |
| C | -1.053067911792 | -2.518265265119 | -1.093531800000 |
| C | 0.072966263599 | 0.085932891857 | 0.013066089720 |
| C | -0.158277305785 | -4.868905721308 | 0.226073669805 |
| C | -0.212446923726 | -0.188005490443 | 2.812389376610 |
| C | 0.412374421419 | -4.738361431050 | 2.779594128750 |
| C | 0.812704358800 | -2.503602008999 | 4.163566749585 |
| H | 3.071160934833 | 2.217172081022 | -0.344445692522 |
| H | 3.740651767411 | -0.736903527218 | -0.407998160868 |
| H | -0.589228372319 | -2.637719595801 | -3.125299036580 |
| H | -1.243591855477 | 1.487843434072 | -0.654910115883 |
| H | -3.070112405855 | -2.430699995748 | -0.963034652450 |
| H | -0.057300531986 | -6.597273671507 | -0.785638713180 |
| H | -2.055676222684 | -0.910019634741 | 3.044227199854 |
| H | 0.974025834216 | -6.503690123269 | 3.740112051570 |
| H | 1.431713876188 | -2.464951064546 | 5.909150484729 |

## S6.3 GFP chromophore HBDI$^-$

The initial structure is obtained starting from the P90 MECI in Ref. 5 but rotating $\phi_P$ by -0.2143° and $\phi_I$ by -85.8572°. $\phi_P$ was rotated from the initial 72.7° to 72.4° and $\phi_I$ was rotated from the initial 17.0° to -68.8°. Only the dihedral angles have been modified, whereas the structure of the 2 rings and the methine bridge have been kept rigid.

Table S9: HBDI$^-$ initial rotated geometry in Bohr, corresponding to $(\phi_P, \phi_R) = (72.283°, -69.338°)$.

| Atom | $x$ | $y$ | $z$ |
|------|-----|-----|-----|
| O | 7.776918745291 | 2.323611600006 | 9.204555374979 |
| O | -0.316144826765 | -4.461287451643 | -1.205721068545 |
| N | 0.084272957587 | -1.095075728299 | -4.003423314879 |
| N | 0.293487044206 | 2.161259791095 | -1.503125191308 |
| C | 6.234066932027 | 1.763049281526 | 7.586684880001 |
| C | 5.279207445979 | -0.800572421409 | 7.342917900429 |
| C | 5.236366469809 | 3.650534445580 | 5.849145476319 |
| C | 3.507734462905 | -1.344098324294 | 5.635763196084 |
| C | 3.386703158239 | 3.058477009156 | 4.236352804112 |
| C | 2.348485142698 | 0.560089439764 | 4.111090219886 |
| C | -0.000000000000 | -0.000000000000 | 2.702372449512 |
| C | 0.000000000000 | 0.000000000000 | 0.000000000000 |
| C | -0.123310731504 | -2.078439309446 | -1.561220337633 |
| C | 0.354697678536 | 1.410410139114 | -3.849773724303 |
| C | 0.196612068388 | -2.600436588752 | -6.255108094215 |
| C | 0.702012236290 | 3.014049280019 | -6.103611154541 |
| H | -1.311084192149 | -2.100296981874 | -7.537904373783 |
| H | 2.211096465899 | 2.294755213707 | -7.281723054175 |
| H | 1.175547633126 | 4.901824539753 | -5.516261203737 |
| H | 6.012296020762 | -2.188866080853 | 8.617178908818 |
| H | 6.044064180588 | 5.500185984979 | 5.968132582479 |
| H | 2.800290110860 | -3.217494907645 | 5.448520092057 |
| H | 2.593532959849 | 4.430645443254 | 2.985870739078 |
| H | -1.572794122697 | 0.983887956163 | 3.526046208710 |
| H | 1.992918293691 | -2.350373263371 | -7.197474982867 |
| H | 0.008250794015 | -4.545316712444 | -5.697112954956 |
| H | -1.000098406522 | 3.075961326636 | -7.239007006443 |



\end{document}